\documentclass[sigconf]{acmart}

\AtBeginDocument{%
  \providecommand\BibTeX{{%
    \normalfont B\kern-0.5em{\scshape i\kern-0.25em b}\kern-0.8em\TeX}}}

\copyrightyear{2025}
\acmYear{2025}
\acmConference[CHI '25 Beyond Glasses: Future Directions for XR Interactions within the Physical World Workshop]{CHI Conference on Human Factors in Computing Systems}{April 26, 2025}{Yokohama, Japan}

\usepackage[utf8]{inputenc}
\usepackage{amsmath}
\usepackage{mathtools}
\usepackage{amsfonts}
\usepackage{xcolor}
\usepackage{listings}
\usepackage[ruled,vlined]{algorithm2e}

\usepackage{float}
\usepackage{makecell} 
\usepackage{ifthen}   
\usepackage{soul}     
\usepackage{csquotes}

\usepackage{pifont}
\renewcommand\footnotetextcopyrightpermission[1]{}

\begin{document}

\title[AfforDance]{AfforDance: Personalized~AR~Dance~Learning~System~with~Visual~Affordance}

\author{Hyunyoung Han}
\orcid{0009-0002-4681-5021}
\affiliation{%
  \institution{Graduate School of Culture Technology, KAIST}
  \city{Daejeon}
  \country{Republic of Korea}
}
\email{hyhan@kaist.ac.kr}

\author{Jongwon Jang}
\affiliation{%
  \institution{School of Computing, KAIST}
  \city{Daejeon}
  \country{Republic of Korea}
}
\email{jangjwhw@kaist.ac.kr}

\author{Kitaeg Shim}
\affiliation{%
  \institution{Graduate School of Metaverse, KAIST}
  \city{Daejeon}
  \country{Republic of Korea}
}
\email{skt0725@kaist.ac.kr}

\author{Sang Ho Yoon}
\orcid{0000-0002-3780-5350}
\affiliation{%
  \institution{Graduate School of Culture Technology, KAIST}
  \city{Daejeon}
  \country{Republic of Korea}
}
\email{sangho@kaist.ac.kr}

\renewcommand{\shortauthors}{Han et al.}

\begin{abstract}
  We propose AfforDance, an augmented reality (AR)-based dance learning system that generates personalized learning content and enhances learning through visual affordances. Our system converts user-selected dance videos into interactive learning experiences by integrating 3D reference avatars, audio synchronization, and adaptive visual cues that guide movement execution. This work contributes to personalized dance education by offering an adaptable, user-centered learning interface.
\end{abstract}



\begin{teaserfigure}
\centerline{\includegraphics[width=0.85\textwidth]{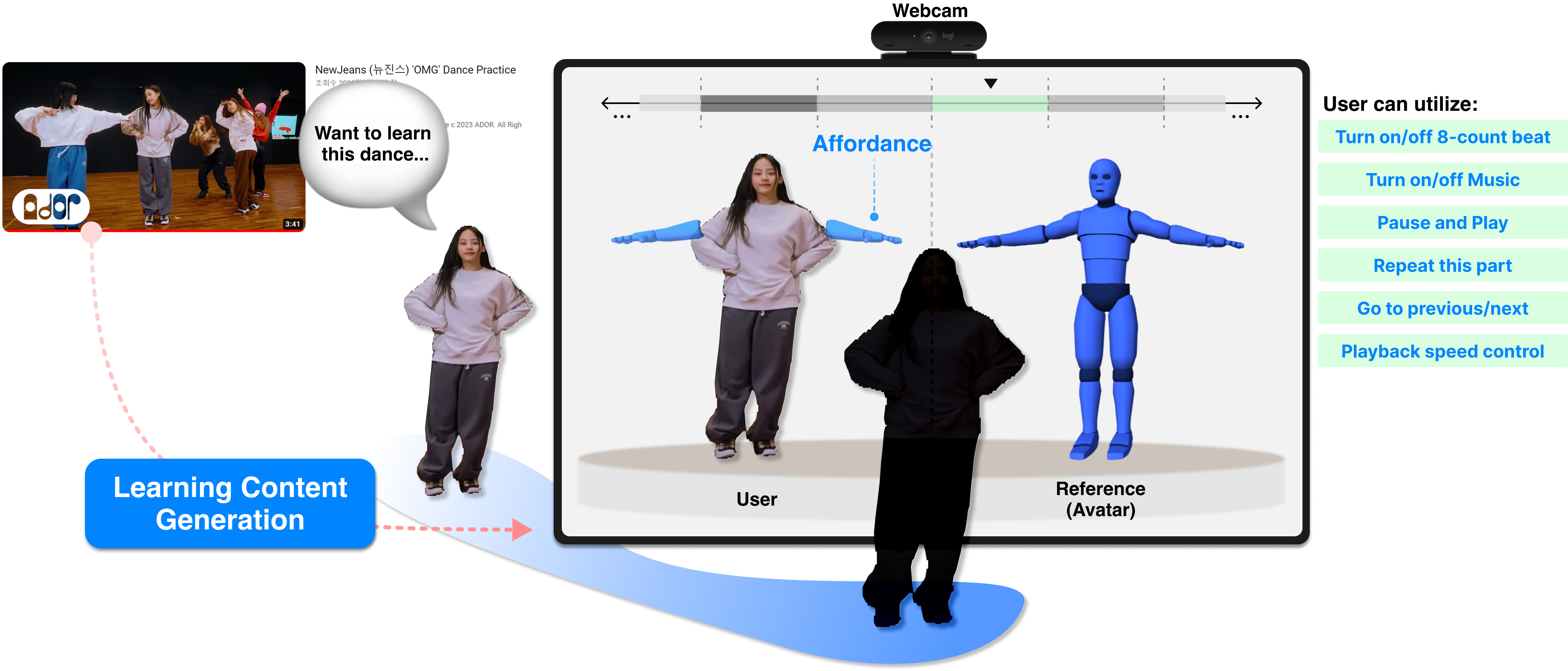}}
  \caption{The proposed system, AfforDance allows users to input a dance video they wish to learn. Our system converts the dance video into learnable content, which users can engage with through the proposed user interface, utilizing an additional webcam and display. The user interface includes core functions such as affordances that enable users to learn detailed movements in depth, as well as various features that facilitate learning at the user's own pace.}
  \Description{This is the teaser figure for the article.}
  \label{fig: teaser}
\end{teaserfigure}


\maketitle

\section{Introduction}

With the rise in accessibility to dance and the proliferation of ``Dance Challenge'' content on video-based social networks like YouTube and TikTok, learners increasingly desire to learn various dance choreographies within short periods~\cite{blanchet2023learnthatdance, wang2022will}. The traditional method of dance learning involves receiving instruction from a teacher, allowing systematic learning but not necessarily offering the dance styles learners wish to acquire. Typically conducted in group settings, it does not accommodate personalized learning based on individual skill levels and requires adherence to fixed times and locations. On the other hand, self-dance learning through online dance videos offers the advantage of learning at one's own pace and choosing desired dances. However, it faces limitations such as time delays in producing learning content and bottlenecks when specific learning materials are unavailable.

Previous studies on dance learning have primarily focused on transferring the learning environment (to the mobile environment such as a tablet~\cite{usui2015learning}, Augmented Reality (AR)~\cite{anderson2013youmove, iqbal2021augmented, bacs2023ar}/Virtual Reality (VR)~\cite{laattala2024anticipatory, chan2010virtual}) to reduce spatiotemporal constraints and enhance learning effectiveness for learners. However, fully immersive HMD-based dance experiences~\cite{laattala2024wave, wu2019vras} pose challenges such as increased physical fatigue due to prolonged headset use and potential spatial disorientation, which can disrupt natural movement flow. These constraints make it difficult for learners to engage in sustained dance practice, necessitating alternative XR-based approaches that balance immersion and usability. The motivation in the learning context lies in whether learners can acquire the dances they desire at a pace suited to their learning capabilities. Therefore, the proactive development of dance learning content and interfaces that reflect user demand and learning receptivity is crucial. This study aims to implement AR-based learning content and interfaces utilizing the characteristics of augmented reality to enhance the freedom of the learning experience~\cite{chang2020applying}. We ultimately aim to \textbf{(1) develop an algorithm for generating dance learning content based on learner preferences}, reflecting the motivation of users who want to learn specific choreographies, and \textbf{(2) propose an AR-based learning interface and visual cues known as affordances} that facilitate effective and suitable dance motion learning tailored to individual learners' capabilities.

\section{Exploratory Study}
\begin{table*}[t]
\centering
  \caption{Demographic and Background Information of Participants in the Contextual Inquiry}
  \label{table1}
    \begin{tabular}{|l|l|l|l|l|l|}
    \multicolumn{1}{|c|}{\textbf{Participants}} & \textbf{Gender} & \textbf{Age} & \textbf{Dance Experiences}  & \textbf{What they learned}                 & \textbf{Amount of learned/initial desired}  \\
    P1 & F & 26 & More then 10 years & LE SSERAFIM - \enquote{Perfect Night} & \textcolor{blue}{8-eight}/8-eight \\
    P2 & M & 26 & Nonexistent        & PSY - \enquote{New Face}              & \textcolor{red}{4-eight}/8-eight  \\
    P3 & F & 26 & Nonexistent        & TWS - \enquote{Plot twist}            & \textcolor{red}{2-eight}/6-eight  
    \end{tabular}
\end{table*}

\subsection{Study Design}
To establish design guidelines for augmenting the dance learning process in an AR environment, we conducted a Contextual Inquiry (CI) to gather user-centered requirements for designing our solution \cite{beyer1999contextual}. We set three goals for conducting the CI:

\begin{itemize}
    \item To deeply understand the culture and context of users’ dance learning through various methods.
    \item To identify the overall experiences (actions, thoughts, emotions) of users with different backgrounds in dance learning.
    \item To identify the major difficulties in dance learning.
\end{itemize}

We conducted the CI in a dance studio at the KAIST campus. The entire session was recorded using a camera and a mobile phone recorder with the participants’ consent. We proceeded with CI involving three participants (2 females and 1 male; mean age: 26) with diverse backgrounds in dance learning as shown in Table \ref{table1}, and they received compensation worth 10,000 KRW. The CI was conducted over 70 minutes following the sequence of \textit{introduction, ice-breaking, observation \& interview,} and \textit{wrap-up}. To elicit the participants’ natural behavior and observe their original actions, we simply asked, \enquote{Please show us what you usually do when you come to self-learn choreography,} without giving any instructions on what song or part they should learn.


\subsection{Findings and Design Goals}
Here, we describe the two major challenges in self-learning from our contextual inquiry and propose three design goals to develop our solution.
\subsubsection{Device Usage Disrupts Learning Flow} 
Participants reported that the size of commonly used device interfaces, such as smartphones and laptops, is unsuitable for learning. P1 and P3 stated, \textit{“It is hard to watch myself in the mirror and the device at the same time.”} Additionally, they experienced discomfort from moving forward to check the video and then back to dance, disrupting their learning process. Therefore, it is necessary to develop a learning environment and interface that reflect the context of movement-centric dance learning while maintaining a natural flow of learning.

\subsubsection{Difficulty in Learning Dance Movements Independently}\label{movement_challenge} 
Participants faced challenges with simultaneous upper and lower body movements (P1-3), segments with numerous actions (P1-3), and synchronizing their movements with the beat at fast tempos (P2, P3). They often forgot the next move (P1-3). Some participants watched dance guide videos for unfamiliar movements but struggled when specific guides were unavailable (P1, P3). Therefore, it is essential to create structured content that allows learners to systematically study their desired dance while enabling them to learn at a pace that matches their individual capabilities.

\subsubsection{Design Goals}
Building on the context of self-learning and our design implications, we propose the following fundamental design goals (DG):

\textbf{DG1: Create Dance Learning Content that Reflects Learner Preferences} We need to develop a dance learning content generation system that reflects learner preferences, enabling them to select and learn the dance they are interested in. This approach enhances learners' motivation to acquire dance choreographies they find intriguing, while simultaneously addressing the lack of learning content mentioned in \ref{movement_challenge}.

\textbf{DG2: Design AR Affordance aiding in the learning of dance movements} Provide optimized visual cues (\enquote{affordances}) using 3D avatars within the AR environment to enable users to effectively learn dance motions. This approach overcomes the aforementioned difficulty of learning movements solely from observing videos when learning independently.

\textbf{DG3: Design the Interface Tailored for Self-Dance Learning} Enhance the self-learning experience by implementing an interface that reflects the actual learning experiences and features used in dance classes. This approach ensures that learners can adapt the learning process to their individual speeds and capacities, making the experience more effective and engaging compared to traditional methods like YouTube videos.

\section{Affordance}
We propose \textbf{AfforDance} (Fig. \ref{fig: teaser}), a system consisting of an AR-based user interface and a learning content generation system. Users can utilize our system with their webcam and display. The process begins with the user providing a dance video they wish to learn. The system then converts this video into learnable content. The user engages with this converted learning content through the proposed user interface. By leveraging affordances, users can learn detailed dance movements effectively, and the interface includes features that enable users to learn at their own pace, similar to actual dance classes.

\subsection{Learning Content Generation} 
\begin{figure}[th]
    \centering
    \includegraphics[width=\linewidth]{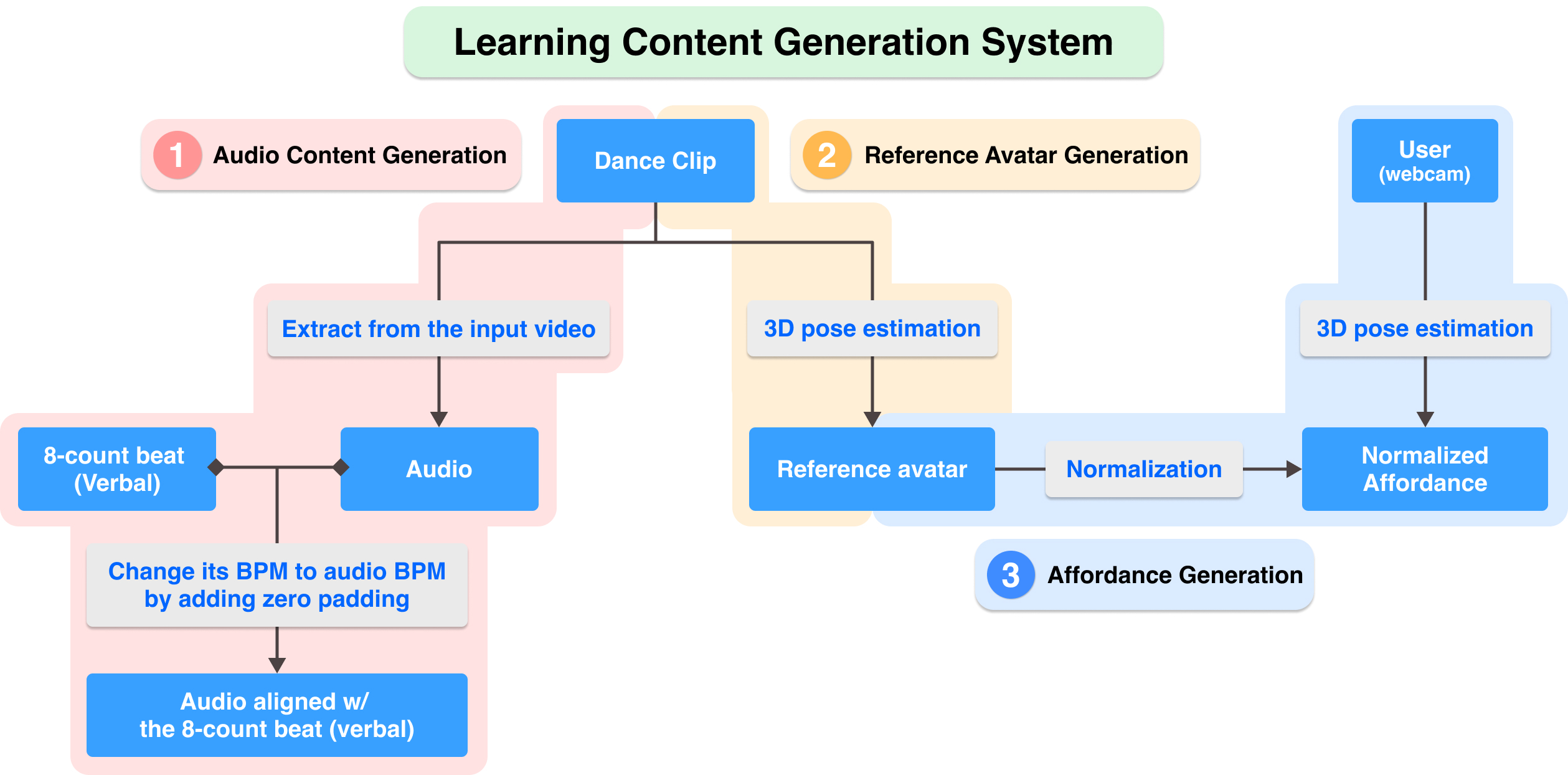}
    \caption{The learning content generation process consists of three stages: (1) Audio content generation, where an 8-count voice beat is added to the extracted audio; (2) Reference avatar generation, creating a 3D avatar from the dance video; and (3) Affordance generation, adjusting affordances to match the user's body size using webcam data.}
    \Description{learning content generation}
    \label{fig:learning content generation}
\end{figure}

The learning content generation system described in Fig.~\ref{fig:learning content generation} consists of three main components: audio content generation, video content generation, and affordance generation, all integrated using the Unity game engine~\cite{unity}.

Dance movements are typically learned in segments known as 8-counts~\cite{dearborn2006dance}. The process of audio content generation involves embedding an 8-count beat into the audio extracted from a dance video. Initially, the system requires the input of a YouTube video ID, the start and end seconds of the desired segment, and the BPM (beats per minute)  of the song. First, the highest-resolution downloadable YouTube stream is selected using \textit{pafy}~\cite{FSF2007}, and audio is extracted with \textit{moviepy}~\cite{Zulko2023}. The 8-count beat’s BPM is adjusted to match the original audio, and both are volume-normalized using \textit{librosa}~\cite{McFee2015}. The adjusted beat is then combined with the original audio, with zero padding for alignment. If needed, manual adjustments are made using Audacity~\cite{audacity}.For reference avatar generation, we extracted body poses from input 2D dance videos and rendered them as meshes to display 3D reference avatars in AR space. We employed the WHAM model~\cite{shin2024wham}, which is a SOTA model in 3D human pose estimation. 

\subsection{Affordance}
\subsubsection{Design and Operation of Affordance}\label{affordance generation} 
\begin{figure}[th]
    \centering
    \includegraphics[width=\linewidth]{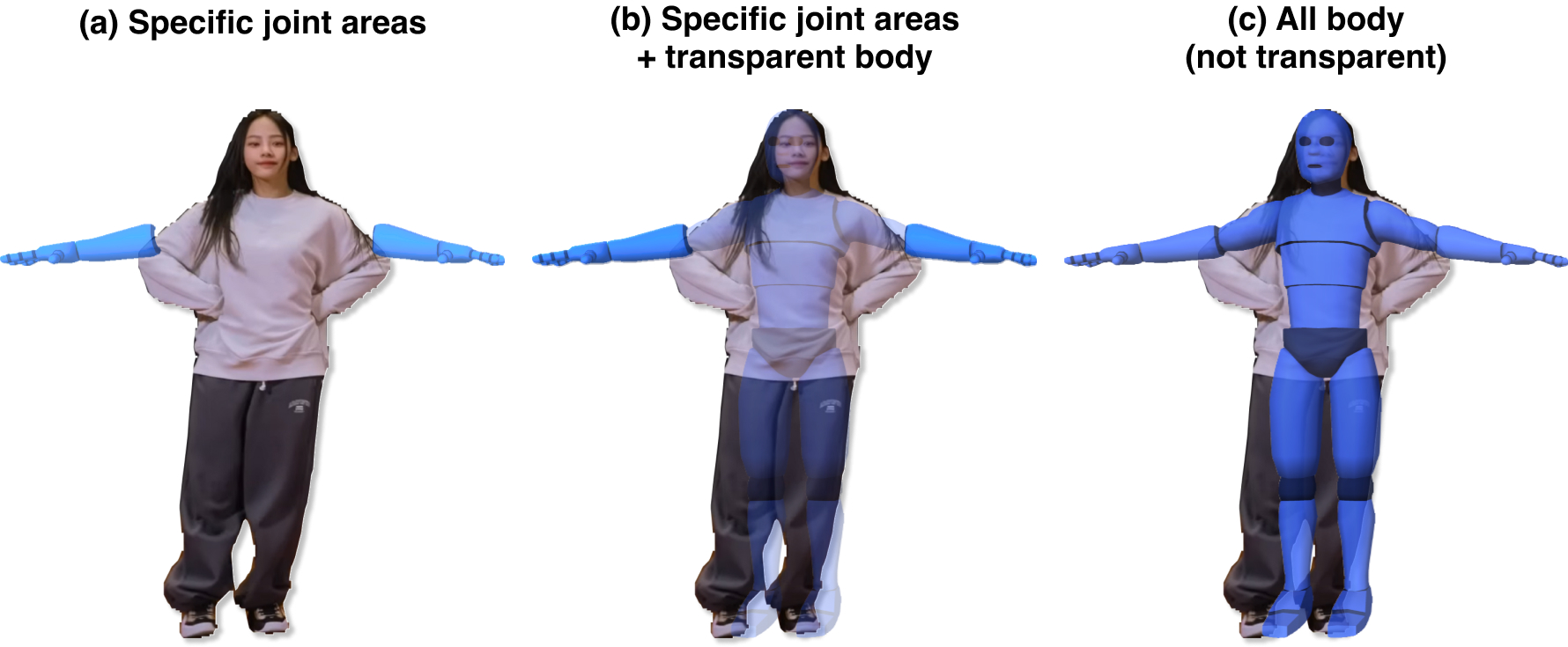}
    \caption{The affordance can be presented as (a) only certain joint areas such as hands and legs, (b) certain joint areas and the rest of the body with transparency, or (c) all body parts without transparency.}
    \Description{affordance type}
    \label{fig:Affordance type}
\end{figure}

Affordance is AR visual cues based on body representation. Users can utilize affordances adjusted to their body size to gain a high level of understanding of the mechanics of dance movements, including body movements and movement range. It is difficult to learn hand and foot movement compared with other parts, prior research~\cite{eaves2011short} demonstrated the effectiveness of using VR feedback on four joints (left/right wrists and ankles) within dance learning. Therefore, affordances are prominently displayed on both hands and feet. Affordances can be applied and displayed in three different modes, as shown in Fig. \ref{fig:Affordance type}.

\subsubsection{Affordance Generation} The affordance generation process consists of raw affordance generation and normalization. The avatar mesh, generated by the WHAM model in PKL format, is first converted to FBX using VIBE~\cite{kocabas2019vibe} and then imported into Blender~\cite{blender}, where affordances are created by separating selected vertex groups into a new mesh. Normalization ensures affordances align with the user’s physical position. Our system employs ThreeDPoseUnityBarracuda~\cite{digital2021} for real-time pose estimation, capturing 24 key body points and kinematic data. To reduce computational overhead, normalization is performed initially rather than every frame. Parameters from the estimated results are applied to the affordance, and the transform is confirmed to ensure proper alignment.

\begin{figure}[tbh]
    \centering
    \includegraphics[width=\linewidth]{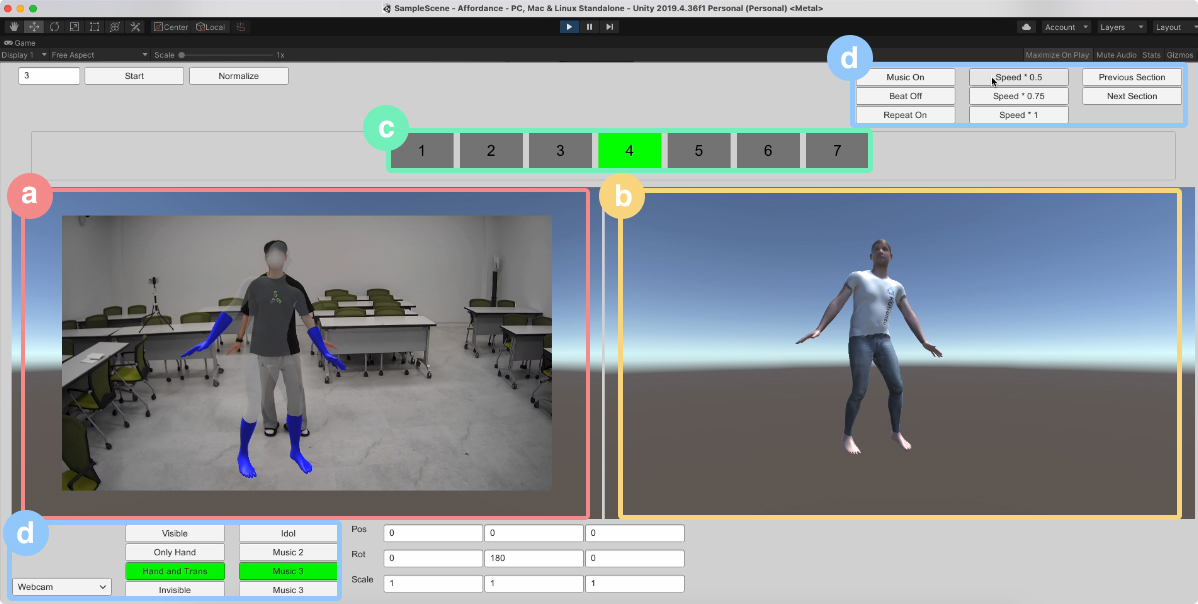}
    \caption{Our user interface consists of four elements: (a) user appearance and affordances, (b) reference avatar, (c) learning progress bar, and (d) learning support features.}
    \Description{User Interface of AfforDance.}
    \label{fig:UI}
\end{figure}

Fig. \ref{fig:UI} illustrates our system's user interface, which consists of several components to support dance learning. The user's webcam-captured appearance and affordances (Fig. \ref{fig:UI}a) are displayed alongside the reference avatar extracted from the dance video (Fig. \ref{fig:UI}b). The timeline interface (Fig. \ref{fig:UI}c) provides an overview of learning sections, while learning support features (Fig. \ref{fig:UI}d) include music/beat/repeat toggles, playback speed control (0.5x, 0.75x, 1x), and section navigation. Users can also select their webcam and affordance type, including an invisible option (Fig. \ref{fig:Affordance type}), enabling a personalized learning experience.

\section{Future Direction for Dance Guidance in XR}

This study proposes an algorithm for generating dance learning content based on learner demand and a dance learning application utilizing AR affordances within a large display. While existing HCI research on augmented dance learning has primarily leveraged HMDs, our study highlights that prolonged HMD use can induce physical fatigue, potentially hindering sustained engagement. To address this, we explore an alternative approach using large displays with AR overlays, allowing learners to receive structured guidance without restricting movement. However, a key limitation of this study is the insufficient discussion on how user input can be effectively integrated into the learning process beyond passive content delivery.

Future work should focus on refining interaction techniques that actively incorporate user feedback into real-time guidance. Instead of relying solely on pre-generated cues, incorporating adaptive motion tracking, gesture-based controls, or interactive feedback loops could enhance the responsiveness of the learning system. Additionally, further exploration is needed to optimize how users interact with AR affordances in large display environments to improve movement fluency and engagement. By addressing these challenges, XR-based dance guidance systems can evolve to provide a more adaptive and user-centered learning experience.

\begin{acks}
    This research is supported by Basic Science Research Program through the National Research Foundation of Korea (NRF) funded by the Ministry of Education (No.2340001991).
\end{acks}

\bibliographystyle{ACM-Reference-Format}
\bibliography{sample-base}

\end{document}